\documentclass[12pt]{article}
\usepackage{epsfig}
\usepackage{graphicx}
\usepackage{amsmath}
\begin{document}
\title{Understanding Health and Disease with\\ Multidimensional Single-Cell Methods}
\author{Juli\'an Candia$^{1,2,3}$, Jayanth R. Banavar$^{1}$ 
and Wolfgang Losert$^{1}$\\
$^1${\small\it Department of Physics, University of Maryland, College Park, MD 20742, USA}\\
$^2${\small\it School of Medicine, University of Maryland, Baltimore, MD 21201, USA}\\
$^3${\small\it IFLYSIB and CONICET, University of La Plata, 1900 La Plata, Argentina} 
}
\maketitle

\begin{abstract}
Current efforts in the biomedical sciences and related interdisciplinary fields are focused on gaining a 
molecular understanding of health and disease, which is a problem of daunting complexity that spans 
many orders of magnitude in characteristic length scales, from small molecules that regulate cell function 
to cell ensembles that form tissues and organs working together as an organism. In order to uncover the molecular nature of the emergent properties of a cell, it is essential to measure multiple cell components simultaneously in the same cell. In turn, cell heterogeneity requires multiple cells to be measured in order to understand health and disease in the organism. 
This review summarizes current efforts towards a data-driven framework that leverages single-cell technologies to build robust signatures of healthy and diseased phenotypes. While some approaches focus on multicolor flow cytometry data and other methods are designed to analyze high-content image-based screens, we emphasize the so-called Supercell/SVM paradigm (recently developed by the authors of this review and collaborators) as a unified framework that captures mesoscopic-scale emergence to build reliable 
phenotypes. Beyond their specific contributions to basic and translational biomedical research, these efforts illustrate, 
from a larger perspective, the powerful synergy 
that might be achieved from bringing together methods and ideas from statistical physics, data mining, 
and mathematics to solve the most pressing problems currently facing the life sciences.
\end{abstract}


\section{Introduction}

Single-cell heterogeneity poses a huge challenge in 
the development and improvement of strategies for the diagnosis and treatment of many diseases. Indeed, it is a well-established fact that cells from the same tissue display significant qualitative and quantitative heterogeneities, even within samples obtained from a single individual.  
This inherent biological diversity has complicated efforts to capture the essence of health and disease in terms 
of characteristic behaviors at the single-cell level and has, 
therefore, limited our ability to fully take advantage of new single cell analysis approaches to improve the current practice of 
personalized medicine. 

For instance, Beckman et al.~\cite{beck12} have very recently assessed the impact of single-cell heterogeneity, as well as that of genetic instability, in the development of effective nonstandard strategies for personalized cancer 
treatment. Manifestations of cell heterogeneity in healthy and diseased 
cell samples have ubiquitously been reported in the growing field of 
single-cell biology, ranging from human pluripotent embryonic stem cell cultures~\cite{deso12,tang12,druk12}
and apoptosis mechanisms in cancer cell lines~\cite{schm12}, to reversible adaptive plasticity in tumors such as human neuroblastoma~\cite{chak12} and pressure-driven shape features of C. elegans embryonic cells~\cite{fuji12}. For recent reviews of the impact of tumor heterogeneity at different levels 
(genetic, epigenetic, the tumor microenvironment, the immune response, and other factors such as diet and the microbiota), 
see Refs.~\cite{meac13,burr13,junt13,beda13}.     

The difficulties of pinpointing specific characteristics of different healthy and diseased cell 
subpopulations prompted the development and refinement of 
experimental techniques that allow multidimensional measurements on single cells, such as e.g. multicolor flow 
cytometry~\cite{perf04,sach05,gonz08}, high performance 
kinetic image cytometry~\cite{ceri12}, and the very recently introduced mass cytometry (CyTOF) 
technique~\cite{orna10,bend11,beno11}.   
Certainly, the improvement of these experimental methods allows one  to probe single cells in 
increasingly high-dimensional parameter spaces, which in turn enhances the resolution to identify and 
focus on specific cell subpopulations. 
As the experimental techniques evolve, however, the pressing need for 
improving our ability to process and analyze ``Big Data" in the life sciences becomes increasingly manifest. In fact, we need unbiased, mathematically robust, scalable methods that allow us to 
identify the key parameters that consistently characterize cell subpopulations across different samples in order to build signatures 
of health and disease across length scales spanning many orders of magnitude~\cite{cand13aps}.  

In this review, we summarize current data-driven efforts that leverage single-cell technologies to build robust 
signatures of healthy and diseased phenotypes. 
We focus on two key types of single-cell datasets, namely 
multicolor flow cytometry, in which each cell is characterized by a set of up to 20 measurements corresponding to scattering and fluorescent emission of light upon stimulation by laser beams, and microscopy via high-content image-based screens, 
in which multiple parameters characterize the shape of each cell, often used in combination with biomarker intensity measurements.      
In Section 2, we discuss the challenges arising from biological complexity, emergent phenomena, and cell heterogeneity. 
In Section 3, we review efforts to build phenotypes based on flow cytometry data analysis techniques. In Section 4, we summarize profiling methods for microscopy image-based screens. 
In Section 5, we present the Supercell/SVM paradigm, which is a general approach for emergent phenotyping that can be applied to different kinds of single-cell datasets, including multicolor 
flow cytometry and cell imaging. Finally, in section 6, we present our concluding remarks. 

\section{From Molecules to Cells to Organisms:\\ Complexity, Emergence, and the Cell\\ Heterogeneity Challenge}
   
Certainly, gaining a molecular understanding of disease is the holy grail of current biomedical research and related interdisciplinary fields. 
The complexity of this problem, however, is daunting; from a physical perspective, this complexity 
arises from the plethora of scales, the great diversity of system components, and the sheer size of biological systems. Indeed, a human being is a hierarchically organized, multiscale system that spans about 10  orders of magnitude in relevant length-scales, from water molecules (0.2 nm), DNA molecules (2.5 nm) and proteins (typically 2-10 nm) to cells (10-100 $\mu$m), tissues, organs, and organ systems. Besides the multi-scale nature of the problem, complexity arises from the high dimensional variety of subtypes (e.g. about 300 human cell types and 25000 protein-coding genes) and the astonishingly large system size (e.g. $3\times 10^9$ DNA base pairs contained in each one of the 10-100 trillion cells of a human being).  

Physics is familiar with the concept of emergence, that is, the collective, large-scale generic behavior of a system that manifests itself independently of small-scale details~\cite{ande72}. In macroscopic systems, thermodynamics and statistical mechanics provide the theoretical framework to understand macroscopic emergence, which explain large-scale phenomena such as ferromagnetism and irreversibility. At mesoscopic scales, however, other kinds of emergent behavior may arise out of as yet unknown physical laws~\cite{laug00}. Collisions of macroscopic particles in low concentration granular suspensions, for instance, tend to disrupt the inherent reversibility of low 
Reynolds number fluid flows~\cite{cort08}, and the onset of irreversibility decreases approximately as the square of particle density. In dense granular systems under cyclic shear, rearrangements in dense configurations are highly collective and known to exhibit reversibility of the neighbor topology in mesoscale trajectories~\cite{slot12}. In biophysics, emergence and complexity are often tackled through multi-scale modeling, which is focused on determining the relevant scales involved in specific processes of interest, as for instance coarse-graining lipid self-assembly dynamics~\cite{klei08,shin12} and actin filament networks~\cite{wilh03,baus06}, as well as using information theory to predict and measure the amount of information transduced by molecular and cellular networks~\cite{levc13}.

\begin{figure}[t!]
\centerline{{\epsfysize=2.2in\epsfbox{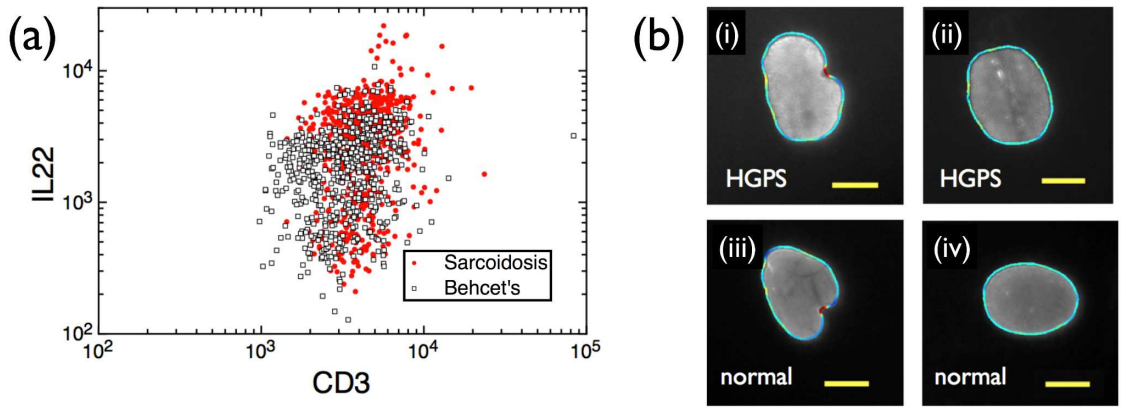}}}
\caption{(a) 2D scatter plot of CD8$^{+}$ T-cell flow cytometry samples from a cohort of 7 patients diagnosed with sarcoidosis and 6 patients diagnosed with Beh\c{c}et's disease, two autoimmune diseases with very similar clinical manifestations. Highly overlapping cell populations (shown as a function of 
the intensity of fluorochrome stains for markers CD3 and IL22) are the hallmark for the phenomenon of cell heterogeneity. (b) Nuclear shapes of diseased ((i) and (ii)) and healthy ((iii) and (iv)) cells can be classified as either blebbed ((i) and (iii)) or non-blebbed ((ii) and (iv)). The diseased nuclei are from Hutchinson-Gilford Progeria Syndrome (HGPS) cell lines. Scale bar: 10~$\mu$. Adapted from Ref.~\protect\cite{cand13}.}
\label{cell_heterogeneity}
\end{figure}

Compounding the multifaceted challenges of emergence and complexity, biological processes are often highly heterogeneous~\cite{alts10}. State-of-the-art technologies, ranging from multicolor flow cytometry and mass cytometry to imaging and single-cell gene expression profiling, are now allowing us to measure multiple properties on single cells. The emerging picture from this growing body of single-cell data shows us that cell heterogeneity is ubiquitous. 
Fig.~\ref{cell_heterogeneity} illustrates the manifestation of cell heterogeneity in different kinds of single-cell datasets. 
In Fig.~\ref{cell_heterogeneity}(a), 100 CD8$^+$ T cells per patient have been randomly sampled from a cohort of 7 patients diagnosed with sarcoidosis and 6 patients diagnosed with Beh\c{c}et's disease, two autoimmune diseases with very similar clinical manifestations. The cells have been stained with 14 fluorochromes; together with forward- and side-scattering, 16 measures have thus been performed on each cell. Despite single-cell flow cytometry data being high dimensional, it is common practice to perform sequential manual partitioning (so-called ``gating") of cell events through visual inspection of scatter plots in 2D. Despite the fact that 
the two markers chosen, CD3 and IL22, are the top markers in a multi-parametric phenotype designed to optimally separate these two diseases 
(see Fig.~\ref{supercell_uveitis_jackknife}(b)), these cell populations are highly overlapping. Similar overlaps are also observed for other cell types and marker pairs. In Fig.~\ref{cell_heterogeneity}(b), images of cell nuclei from a cell line classified as Hutchinson-Gilford Progeria Syndrome (HGPS), a premature aging disorder, are compared to healthy cells. HGPS is known to be caused by mutant 
lamin A, which affects the nuclear scaffolding; the phenotypic hallmark of HGPS is nuclear blebbing~\cite{dris12}. Due to cell heterogeneity, however, we observe both blebbed (i) and non-blebbed (ii) diseased cells, and similarly, blebbed (iii) and non-blebbed (iv) normal cells. 

There are multiple potential mechanisms for the heterogeneity in cell shape.  In addition to the heterogeneity in protein expression levels, 
molecular mechanisms for intracellular transport are also known to be highly heterogeneous~\cite{hoef13,bres13}. 
These transport mechanisms can be broadly classified into two categories: passive diffusion and motor-driven active transport.  
By performing extensive single-tracking microscopy of endogenous lipid granules in living fission yeast cells, it has 
been shown that the granules exhibit a crossover behavior from continuous time random walk subdiffusion at short time scales to subdiffusive 
fractional Brownian motion at longer times~\cite{jeon11}. The associated phenomenon of weak ergodicity breaking leads to the observed crossover in the measured mean squared displacements~\cite{jeon11}. Very recently, other scenarios based on heterogeneous diffusion, in which the diffusion 
coefficient depends on the particle position within the cell, have been explored as well also showing that they lead to sub- and super-diffusive 
regimes with weak ergodicity breaking~\cite{cher13a,cher13b,cher13c}. In the case of active transport, a variety of stochastic approaches 
have been proposed to explain in vivo intracellular observations of tracked biomolecules and microbeads including random walk models, 
quasi-steady-state reduction methods, exclusion processes, random intermittent search processes, Brownian ratchets, along with mean-field approximations (see~\cite{hoef13,bres13} and references therein).

The interplay of complexity, emergence, and cell heterogeneity poses a huge challenge to build reliable molecular phenotypes 
of disease. In the following Sections, we briefly summarize current data-driven efforts based on flow cytometry and cell imaging, although some of those approaches may also be applied to other kinds of state-of-the-art and forthcoming single-cell techniques, such as mass cytometry CyTOF), single-cell gene expression, and single-cell full genome sequencing technologies. 

\section{Single-cell-based Phenotyping using\\ Multicolor Flow Cytometry Data}

Flow cytometry is used routinely both in research labs to investigate cell structure and function as well as in clinical labs to diagnose disease, 
assess its progression and its response to therapy. The flow cytometry technique is based on fluorochrome-conjugated antibodies that attach to cell surface proteins and 
intracellular molecules. Laser beams and detectors are used to measure the forward-scattered light (which provides a measurement of the size of the cell), the side-scattered light (which is used to determine the internal structures within the cell) and the emitted light from each type of fluorochrome (which is proportional to the corresponding antigen density). By using compensation techniques to resolve the overlaps of absorption and emission spectra, 
up to 20 different properties can be determined on a cell-by-cell basis~\cite{baum00,lugl10}.  

Although single-cell flow cytometry data is high dimensional, practitioners usually perform sequential manual gating of cell events through visual inspection of scatter plots in 2D, a procedure aided by a number of analysis software programs such as FlowJo~\cite{flowjo} and FCS Express~\cite{fcsexpress}. Over the past few years, new computational methods have been developed in order to overcome the serious limitations in manual gating-based analysis~\cite{hahn09,naim10,qian10,agha12a,qiu12,ge12,agha12b,zare12}. All these methods share a common rationale, which is first to define a procedure to extract characteristic features from each sample, and then to perform machine-learning-based classification of different sample types. A comparative assessment of the performance and accuracy of different methods is summarized in Ref.~\cite{agha13}. Typically, machine-learning instance space is defined by statistical population measures (such as mean, standard deviation and higher moments), histogram classification (through binning into low-dimensional measurement subsets), clustering of events (e.g. k-means or spectral clustering) 
or mixture models (Gaussian, nonparametric Bayesian, skew-t, etc). Since flow cytometry is high-throughput, with samples typically consisting of hundreds of thousands and even millions of cells, some methods incorporate down-sampling preprocessing steps. In particular, density-based down-sampling methods are designed to filter out denser regions, so that small subpopulations of cells are more significantly represented in the down-sampled datasets than in the original data. 

\begin{figure}[t!]
\centerline{{\epsfysize=3.5in\epsfbox{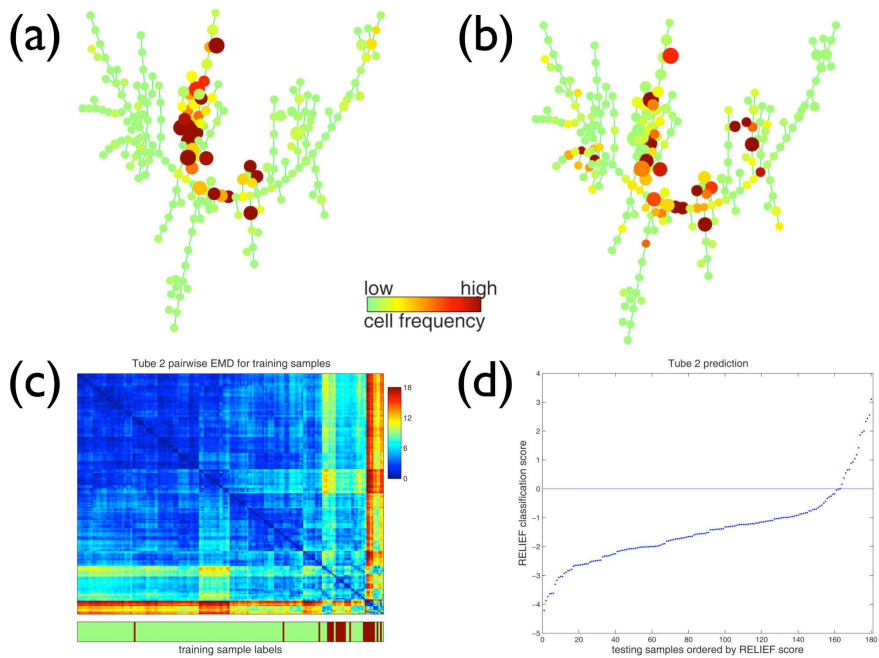}}}
\caption{Cell distribution of different cell subpopulations in the SPADE tree for (a) one healthy donor and (b) one AML patient. (c) Pairwise EMD distance between training samples, where the order of the samples in the heat-map was organized by hierarchical clustering. The bottom panel shows the class label of each training sample, green for healthy and red for diseased. (d) Relief scores to predict test samples. Adapted 
from Ref.~\protect\cite{qiu12}.}
\label{spade}
\end{figure}

For the purpose of illustration, we summarize here one particular approach that has successfully been applied to classifying acute myeloid leukemia (AML) positive patients and healthy donors using flow cytometry data~\cite{qiu12}, which is based 
on the so-called SPADE algorithm~\cite{qiu11} that has also proved useful in cell subpopulation analysis of mass-spectroscopy-based cytometry~\cite{bend11}. First, SPADE is applied to perform the extraction of features based on the relative size of subpopulations. SPADE consists of four computational modules: density-dependent down-sampling, agglomerative clustering, minimum-spanning tree construction, and up-sampling. Using data from all samples (both diseased patients and healthy donors), a tree formed by 
150 clusters was obtained.  It is assumed that groups of clusters roughly correspond to different cell subpopulations. Second, the distribution of cells across the 150 clusters is computed for each sample. Fig.~\ref{spade}(a) shows the cell distribution across the SPADE tree for a healthy donor, while Fig.~\ref{spade}(b) displays the cell distribution for a diseased patient. Third, the pairwise Earth Mover's Distance (EMD) between cell distributions is calculated. EMD measures the minimum effort needed to make one distribution the same as the other by moving cells, which can be calculated as a constrained linear programming problem~\cite{rubn00}. Pairwise EMDs for all training samples (156 healthy donors and 23 diseased patients) are shown in Fig.~\ref{spade}(c), where the order of the samples in the heat-map was organized by hierarchical clustering. The bottom panel shows the class label of 
each training sample, green for healthy and red for diseased. 
Finally, test samples were classified based on their Relief scores. 
Relief is a nearest neighbor based classifier~\cite{kira92}; the Relief score for one test sample is defined as the distance (which, in this case, is determined using the EMD metric) between the test sample and the nearest normal sample minus the distance from the test sample to the nearest AML sample. 
Therefore, test samples with negative Relief scores are classified as normal, whereas samples with positive Relief scores are classified as diseased. Fig.~\ref{spade}(d) shows blind test results for 180 samples ordered from left to right by increasing Relief scores. As it turns out, all the samples with positive Relief scores have been correctly classified as diseased. This procedure can be straightforwardly extended to the case in which multiple test tubes are used to measure different sets of fluorochromes~\cite{qiu12}.

\section{Single-cell-based Phenotyping using\\ Microscopy Images}

The field of biological image analysis has seen rapid progress with the advent of automated image acquisition systems 
that enable new types of microscopy experiments.  A variety of imaging techniques yield high-content, high-throughput 
images from fluorescence microscopy, electron microscopy, bright-field microscopy, differential-interference-contrast 
microscopy, as well as from multi-spectral and multi-dimensional microscopy. 
Yet, the bottleneck for progress in this field appears to lie in the efficacy of computer vision, image analysis, and 
pattern recognition methods~\cite{peng08}. 

\begin{figure}[t!]
\centerline{{\epsfysize=1.7in\epsfbox{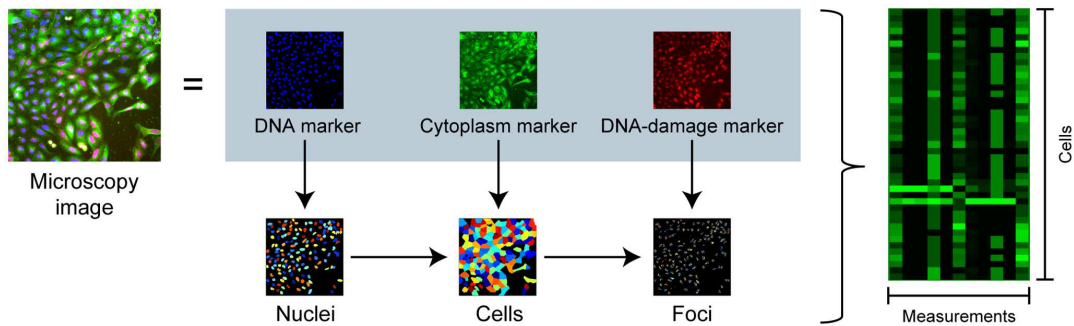}}}
\caption{Multiparameter automated feature extraction from nuclei imaging. In this example, after variations in illumination and staining are corrected, nuclei are identified by thresholding and then used as seeds to identify cell edges. Finally, markers are 
used to identify DNA-damage foci. Multiparametric quantitative single-cell features may then result from transformations of the image-based measurements. From Ref.~\protect\cite{ljos09}.}
\label{image_parameterization}
\end{figure}

The analysis of digital microscopy images usually requires identifying so-called ``regions of interest" (ROIs), e.g. cells, 
cell nuclei or organelles that represent target objects within the images. The process of ROI identification is still sometimes 
performed manually; this approach, however, is labor-intensive and impractical for the analysis of large-scale screens, 
and is prone to introducing procedural biases and inconsistencies. In computer-aided 
image analysis, the first step is usually to determine ROIs and disentangle foreground from background pixels, a process known 
as segmentation, although pattern recognition can also be used to process 
whole images or image tiles on a grid without identifying ROIs~\cite{sham09,raja12}. 
Fig.~\ref{image_parameterization} illustrates the process of ROI definition and extraction in a fluorescence-microscopy image 
of cultured cells from screens. In this example, the goal is to segment individual nuclei and individual DNA-damage-induced foci, 
as well as to segment the cells to characterize their individual morphology. To this end, a set of three fluorescent stains (DNA, 
cytoplasm, and DNA-damage markers) are used and recorded in each image in different channels. 
Some implementations of segmentation algorithms are designed for a very specific type of object (e.g. $3-D$ images of neurons~\cite{peng10}) and hence do not require detailed fitting of parameters, while other tools are of a more general scope and need more 
intensive tuning to target the desired ROIs. Widely used methods include global thresholding~\cite{otsu79,saho88} with or without bias correction~\cite{lind01}, local or adaptive thresholding~\cite{solo99}, segmentation based on texture, local intensity variations or other features~\cite{baat06}, watershed algorithms~\cite{meye90,vinc91,lind93,roer01}, model-based segmentation~\cite{cong00}, level-set approaches~\cite{duan08}, contour methods~\cite{vrom06,dris12b}, and automatic edge detection~\cite{li08}. 

\begin{figure}[t!]
\centerline{{\epsfysize=3.6in\epsfbox{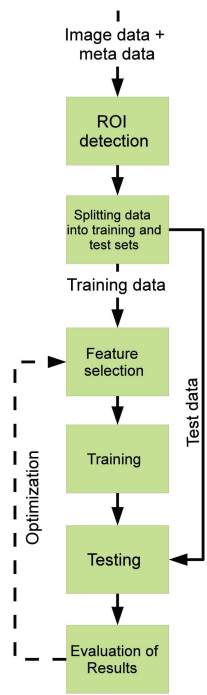}}}
\caption{Process of pattern recognition in biological image analysis: identification of Regions of Interest, extraction of image content descriptors, feature selection and classification, testing on new instances (or, alternatively, leave-one-out cross-validation) and evaluation of results. From Ref.~\protect\cite{sham10}.}
\label{image_analysis}
\end{figure}

After the identification of Regions of Interest, the process continues by extracting image content descriptors, see Fig.~\ref{image_analysis}. Feature extraction algorithms read image pixels from foreground ROIs and output a variety of numerical image features that characterize the assay, including the number of objects per image (e.g. the number of DNA-damage-induced foci 
per cell in the example from Fig.~\ref{image_parameterization}), the intensity of a marker as reflected by pixel intensity 
in a given channel (e.g. the total intensity of DNA marker in the nucleus can be used to identify cell-cycle phases), 
texture descriptors (such as statistical, structural, and spectral texture measures), and shape descriptors that enable 
morphological phenotyping (e.g. Zernicke shape features~\cite{bola98} and Hu moments~\cite{hu62}). 
In some applications, image features are extracted from arbitrarily defined tiles that do not necessarily correlate with 
actual cellular or subcellular features, which is the kind of approach used in algorithms such as PhenoRipper~\cite{raja12} and 
WND-CHARM~\cite{orlo08}. 

Finally, phenotyping healthy vs diseased samples, or chemically and genetically perturbed samples, is usually achieved by 
implementing a machine learning classifier. As shown schematically in Fig.~\ref{image_analysis}, this process is iterative. 
The initial set of features usually ranges between few tens and one hundred, although it is assumed that many of them are 
irrelevant or redundant. Two general approaches to feature selection are known as filters and wrappers~\cite{guyo06}. 
Filtering methods are based on statistical measures before the classifiers are applied. For instance, Fisher scores, which 
measure inter-class variance relative to intra-class variance, can be used to weight features according to their classification 
discriminative power~\cite{orlo08}. Another filtering approach, which aims at avoiding the bias due to 
highly correlated features, is 
the so-called minimum redundancy maximum relevance (mRMR) algorithm~\cite{ding05}. Other filtering methods are based on remapping 
the original feature space into lower dimensional ones, including linear approaches such as singular value decomposition~\cite{holt00,alte00,holt01} and principal component analysis~\cite{rayc00,wall03}, as well as non-linear ones such as 
self-organizing maps~\cite{tama99,koho01}, isomap~\cite{tene00}, local linear embedding~\cite{rowe00} and multidimensional 
scaling~\cite{borg97,trei10}. Wrapping approaches, in turn, select features based on their actual performance in the classifier 
(see e.g.~\cite{guyo06,mark05,beck09}); they oftentimes provide better feature selection and greater classification accuracy 
than filtering, although in most cases wrapping is significantly slower because it relies on running many iterations with different 
feature sets (as indicated by the dashed upwards arrow in Fig.~\ref{image_analysis}). 

A number of proprietary and open-source software tools are available to perform specific image analysis tasks such as segmentation, 
feature selection and classification. Moreover, some packages integrate all the tasks needed along the image analysis workflow. 
Software packages such as WND-CHARM~\cite{orlo08}, CellProfiler~\cite{carp06}, ImageJ~\cite{abra04}, PSLID~\cite{bola01}, Ilastik~\cite{kres11}, and BIOCAT~\cite{zhou13} offer diverse capabilities in terms of graphic user interface, $3D-$image analysis, machine learning classifiers, extensible algorithm plugins, ROI detection, automatic comparison among algorithms, commercial software requirements, and OS platform requirements.  

\section{Emergent Phenotyping via the\\ Supercell/SVM Paradigm}

Flow-cytometry-based phenotyping methods (see Sect. 3 and references therein) usually rely on the fact that typical cytometry experiments are able to perform multidimensional single-cell measurements on tens and hundreds of thousands, if not even millions, of cells. Therefore, single-cell-level heterogeneities such as those described in Sect. 2 are averaged out by means of large population averages. High-content image-based phenotyping methods (see Sect. 4 and references therein), in turn, usually focus on pattern recognition and classification of regions of interest, but typically disregard cell-to-cell heterogeneities within a given phenotype. 

Heterogeneity from cell to cell, however, is now recognized in many cases not as noise, but as an important feature of many living systems that enables adaptation to changing environmental conditions for a heterogeneous ensemble, from bacterial persistence under antibiotic treatment~\cite{bala04,kuss05} to plasticity and functional diversity in the human brain~\cite{mcco13}.  Single-cell heterogeneity is often encountered in biomedical basic and translational research, as well as in the clinical realm, and leads to particular challenges for studies that are based on a limited number of single cells.  There are two overall constraints that limit the number of analyzed cells. In some diseases, the number of cells that can be collected is limited; for instance, diagnosis of intraocular lymphoma is performed through difficult specimen collection procedures that yield very low numbers of collected cells, typically in the range of ten to a hundred cells~\cite{gonz07}.  In the field of stem cell research, identifying stem cell phenotypes through flow cytometry cell sorting often relies on extremely rare subpopulations. For instance, long-term hematopoietic stem cells (LT-HSCs) identified via immunophenotypes such as Lin$^-$Kit$^+$Sca$^+$CD34$^{lo}$Flt3$^-$~\cite{chri01} and SLAM~\cite{kiel01} represent only about $0.0075\%$ of the cells from whole bone marrow specimens; thus, more than a million whole bone marrow cells need to be extracted, stained with multiple fluorochromes and sorted in order to yield about one hundred LT-HSCs. In early embryonic development studies, e.g. at the stage of mouse early development 
when the founder population of germline cells have just emerged, there are just about 30 primordial germ cells in the embryo~\cite{sait02}. 
The second constraint is due to the analysis approach, where more sophisticated analysis with a larger number of parameters (i.e. higher dimensional data) usually are accompanied by lower numbers of cells analyzed.  Image based screens are typically based on tens to hundreds of cells, with higher dimensional analyses (e.g. 3D cell shapes, multispectral images) often based on fewer cells than simpler lower dimensional datasets. State-of-the-art single-cell genomics technologies are an extreme example of this trend.  It is now possible to measure the expression level of all $20,000+$ genes in a single cell~\cite{tang09}, but the number of cells for which all genes can be measured is limited by both cost and instrument capacity. For these new high dimensional data with limited numbers of datapoints, data analysis methods that rely on high-dimensional clustering procedures, Gaussian mixture approximations, etc may be expected to fail. 

Therefore, it is of paramount importance to address 
the problem of phenotypic classification when single cells are highly heterogeneous and the number of cells available is small. 
Within this context, we have recently proposed the so-called ``Supercell/SVM Paradigm"~\cite{cand13} as a general method for single-cell 
phenotyping that focuses on mesoscopic-level emergent properties of groups of cells. 
The key contribution of this method is to provide a quantitative assessment of the critical sample size and number of simultaneous single-cell measurements needed to identify a phenotype with strong predictive power. 

\begin{figure}[t!]
\centerline{{\epsfysize=2.9in\epsfbox{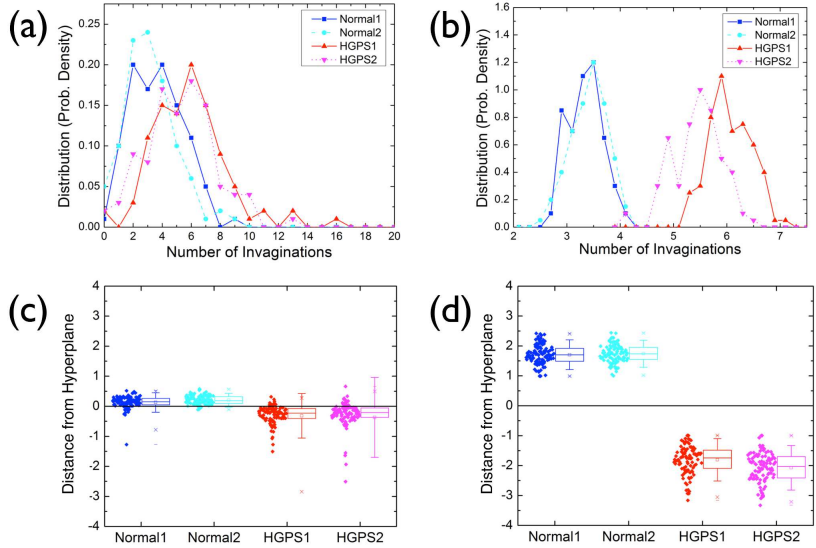}}}
\caption{Analysis of nuclear shapes from images obtained from two normal and two HGPS cell lines. 
Probability density distributions for the number of invaginations of the nuclear boundary corresponding to: (a) single cells 
and (b) supercells of size $N=30$. Distances to the SVM boundary by applying machine learning to (c) single cells and (d) supercells of size $N=30$, where 15 parameters per nucleus are used to characterize shape features and concentration levels of lamin A/C. Adapted from Ref.~\protect\cite{cand13}.}
\label{supercell_progeria}
\end{figure}

In order to capture emergent behavior, a ``supercell of size $N$" is defined as the average of the individual measurement vectors 
of a group of $N$ randomly chosen cells. By repeatedly taking different random subsets of $N$ cells, ``supercell samples" can be built 
out of the original single-cell datasets; on these samples, ``supercell statistics" can be readily computed. This procedure is illustrated in Fig.~\ref{supercell_progeria}(a)-(b), where probability density distributions are shown for one shape parameter (the number of invaginations of the
nuclear boundary) obtained from images from two healthy and two HGPS cell lines. As mentioned above, HGPS is a premature aging disorder caused by mutant lamin A that affects the nuclear scaffolding; although HGPS is characterized by nuclear blebbing, phenotypes at the single-cell 
level are highly heterogeneous (recall Fig.~\ref{cell_heterogeneity}(b)). 
The single-cell distributions in Fig.~\ref{supercell_progeria}(a) are highly overlapping, reflecting the fact that, based on individual cells, one is not able to distinguish healthy cells from diseased ones. After applying the cell averaging procedure (using $N=30$ randomly selected cells to generate each supercell), the resulting 
distributions do not show any significant overlap between healthy and diseased samples (see Fig.~\ref{supercell_progeria}(b)). 
Indeed, the removal of distribution overlaps is a manifestation of the central limit theorem, by which distributions of supercells 
of size $N$ are expected to become narrower by $\sim 1/{\sqrt N}$. 
After cell averaging, machine learning is used to learn what combination of parameters best distinguishes healthy from diseased cells. 
The method implemented in Ref.~\cite{cand13} is a Support Vector Machine (SVM) with a linear kernel, which is less prone to overfitting issues than nonlinear mappings~\cite{cris00,witt11,tarc07}. Moreover, the components of the vector normal to the boundary hyperplane can be  
straightforwardly interpreted as amplitudes that determine the relative significance of the measured parameters in achieving class separation. Healthy and HGPS nuclear shapes were characterized by 12 parameters including minor/major axis length, perimeter, area, tortuosity, mean and standard deviation of the curvature, and the number of invaginations of the nuclear boundary. Moreover, 3 fluorescence intensity measurements associated with the concentration of lamin A/C were measured on each nucleus, thus contributing 3 additional parameters. However, for single cells, even 15 parameters do not suffice to learn the distinction between healthy and diseased individual cells. Indeed, Fig.~\ref{supercell_progeria}(c) shows the distance from each cell to the SVM boundary, where positive (negative) distances correspond to the boundary side identified with the healthy (diseased) class: it is observed that some cells from the healthy cell lines are classified as diseased, and vice versa. Instead, by applying machine learning to supercell distributions, $100\%$ classification accuracy is achieved, as shown in Fig.~\ref{supercell_progeria}(d).

\begin{figure}[t!]
\centerline{{\epsfysize=1.7in\epsfbox{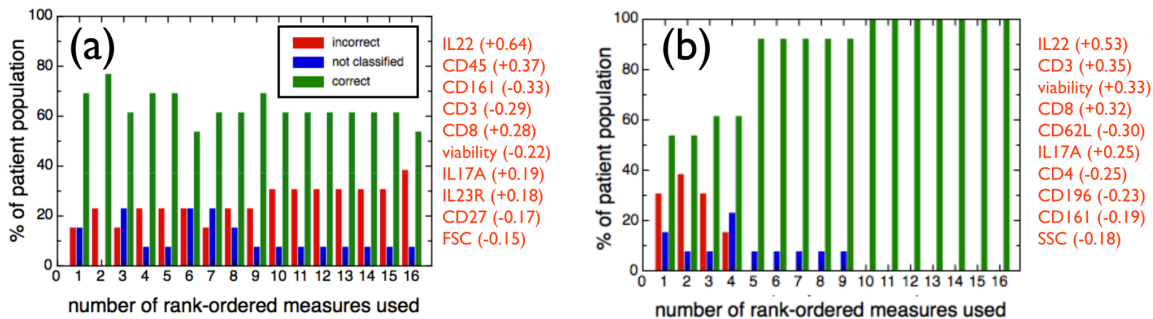}}}
\caption{Jackknife results for sarcoidosis vs Beh\c{c}et's disease using supercells of size $N=500$, where each patient is represented 
by a cloud of 100 supercells, as a function of the number of rank-ordered measures used: (a) all cells; (b) CD8$^+$ T cells. The bars show 
percentages of correct (green), unclassified (blue) and incorrect (red) predictions. To the right of each panel, the list of the top-10 rank-ordered 
measures is shown. Adapted from Ref.~\protect\cite{cand13}.}
\label{supercell_uveitis_jackknife}
\end{figure}

In the analysis of HGPS through nuclear shape measurements, the Supercell/SVM method determined that about 30 cells are needed to classify 
samples as healthy or diseased, in agreement with usual lab procedures. Besides providing validation of current wet-bench practice, the supercell method lays out a straightforward framework towards determining 
how many cells should be measured in novel scenarios (e.g. drug testing, gene-knockout phenotypes, etc) where carrying over ``tried and tested" rule-of-thumb lab procedures may be inappropriate. 

Let us consider next a more challenging problem, namely that of building flow-cytometry-based molecular phenotypes for two non-infectious uveitides (the ocular manifestations of sarcoidosis and Beh\c{c}et's disease), which are very difficult 
to diagnose. By performing 2 scattering and 14 fluorescent measurements on each cell, samples from 7 sarcoidosis and 6 
Beh\c{c}et's patients were measured. Since the cohort was small, prediction testing was carried out by a jackknife (leave-one-out) cross-validation procedure~\cite{shao95}. 
The SVM boundary allows one to rank-order the 16 measures from most to least significant, according to the components of the vector normal to 
the hyperplane that separates the two diseases. Thus, one can selectively remove the least significant measurements from the list and explore 
the minimal number of measures needed to correctly predict the class of all (or at least most of) the samples.  
Fig.~\ref{supercell_uveitis_jackknife} shows jackknife results for supercells of size $N=500$, where each patient is represented 
by a cloud of 100 supercells, as a function of the number of rank-ordered measures used. The list of the top 10 measures is shown to the right 
of each panel. Fig.~\ref{supercell_uveitis_jackknife}(a) shows jackknife results based on all cells, while Fig.~\ref{supercell_uveitis_jackknife}(b) displays results based on CD8$^+$ T cells, a subpopulation that can be determined by manual gating 
(CD3$^+$viab$^-$CD8$^+$CD4$^-$) and typically represents about $5\%$ of the peripheral blood sample. Since each patient is represented by a cloud of supercells, a prediction was made only when more than $95\%$ of those supercells lie on any one side of the SVM boundary. Correct predictions are shown by green bars, incorrect predictions by red bars, while unclassified samples are shown in blue. While predictions based on all cells are very poor, for CD8$^+$ T cells no failed predictions are incurred when five or more measures are used. Therefore, the top five measures listed in Fig.~\ref{supercell_uveitis_jackknife}(b), if linearly combined as $29\%$~IL22 + $19\%$~CD3 + $18\%$~viab + $17.5\%$~CD8 - $16.5\%$~CD62L, can be used on CD8$^+$ T cells as molecular phenotypes that distinguish the two diseases.  

\begin{figure}[t!]
\centerline{{\epsfysize=1.7in\epsfbox{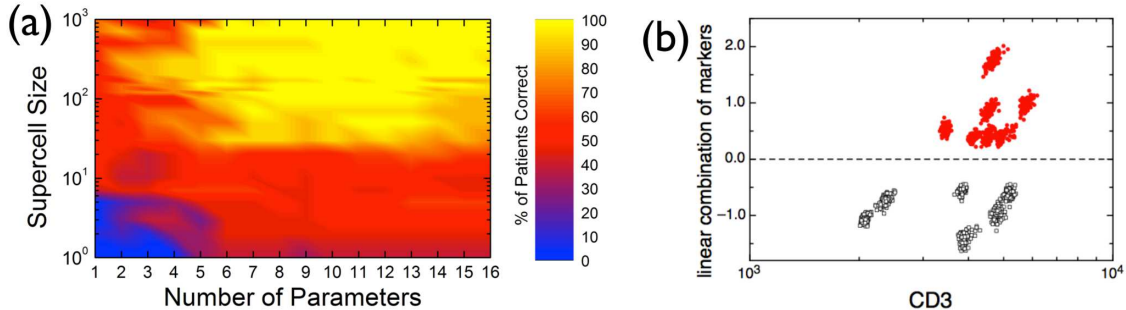}}}
\caption{(a) Percentage of correctly classified CD8$^+$ T supercells as a function of the supercell size 
and the number of measures used. (b) Linear combination of the top 5 markers IL22, CD3, viability, CD8 and CD62L, as a function of CD3, for supercells averaged over 500 randomly chosen CD8$^+$ T cells. Adapted from Ref.~\protect\cite{cand13}.}
\label{supercell_uveitis_separation}
\end{figure}

In Fig.~\ref{supercell_uveitis_separation}(a), the percentage of correctly classified CD8$^+$ T supercells is shown as a function of the supercell size and the number of measures used. It turns out that, by using the top five markers, slightly less than 100 cells are sufficient for reliable prediction. Increasing the number of markers or averaging over more cells, however, do not significantly change the reliability of the classification. The Supercell/SVM framework, thus, allows one to determine which and how many measures, as well as how many cells, need to be measured. Because adding markers and measuring larger samples are expensive and time consuming (and, as mentioned above, might be strongly constrained by experimental and clinical considerations), the ability to determine the minimal setup required to build reliable phenotypes is of paramount importance. Finally, Fig.~\ref{supercell_uveitis_separation}(b) presents a novel visualization based on the usual $2D$ scatter plots used in flow cytometry to find cell subpopulations (recall Fig.~\ref{cell_heterogeneity}(a)). By allowing linear combinations of markers in the vertical axis, a clear separation between disease classes can be readily observed. 

\section{Conclusions}

By complementing more traditional approaches based on computational (and, to a lesser extent, analytical) modeling, data-driven research is 
opening up new avenues of investigation fueled by the advent of powerful high-throughput biomedical technologies.  
Within that context, this review focuses on current efforts towards developing quantitative, unbiased, 
mathematically robust methods to learn healthy and diseased phenotypes from multidimensional, single-cell biomedical data. 

In order to understand health and disease with multidimensional single-cell methods, one needs to overcome the challenges arising from 
the interplay of complexity, emergence, and cell heterogeneity. Analysis methods based on multicolor flow cytometry 
usually rely on rich datasets typically comprising hundreds of thousands (sometimes even millions) of cells. Under such high-throughput 
conditions, single-cell-level heterogeneities are often simply averaged out, thus allowing population-level signatures to emerge. 
A number of machine-learning-based classification methods have been developed recently; they define different kinds of instance spaces, 
on which machine learning classifiers are trained and then used to predict on new samples. 
For a comparative assessment of the performance and accuracy of different automated flow cytometry data analysis techniques, see Ref.~\cite{agha13}. 

High-content image-based phenotyping methods, in turn, usually focus on pattern recognition and classification of regions of interest. In some cases, however, image classification and phenotyping relies on pixel tiles that do not correlate with biologically relevant measures such as cell morphology and biomarker intensity. Moreover, pattern recognition methods are focused on 
pattern classification as opposed to phenotype classification, in the sense that they 
typically disregard cell-to-cell heterogeneities within a given phenotype. For reviews of image-based approaches, see Refs.~\cite{peng08,ljos09,sham10}.

In this review, we emphasize the so-called Supercell/SVM paradigm (recently developed by the authors of this review alongside collaborators, see Ref.~\cite{cand13}) as a unified framework that captures mesoscopic-scale emergence to build reliable phenotypes. Supercells are multidimensional 
objects that represent the collective behavior of groups of cells; within this approach, supercells represent the building blocks of 
healthy and diseased phenotypes. From a conceptual standpoint, this approach naturally incorporates emergent behavior and thus 
cell heterogeneity, usually regarded as a roadblock in the pursuit of characterizing single-cell-level behavior, 
becomes the fundamental conceptual unit to identify collective phenotypes. From a practical perspective, furthermore, the Supercell/SVM framework provides a quantitative assessment of the critical sample size and the number of simultaneous single-cell measurements needed to build a phenotype, which is a key piece of information given the fact that, in many single-cell applications, the number of measured cells and the number of measurements per cell are severely limited due to a variety of constraints, such as experimental costs, technological capabilities, specimen collection procedures, the availability of specialized personnel, and others. 

Finally, let us point out the importance of interdisciplinary exchanges of methods and ideas such as these, in which tools and concepts 
from statistical physics, data mining, and mathematics are successfully applied to the most pressing problems in the life sciences. 
We hope that the powerful synergy of collaborative efforts across disciplines grows ever stronger and bears fruit by improving diagnosis, prognosis, and decision making in the clinical realm.

\section*{Acknowledgments}
We acknowledge our coauthors A. Biancotto, K. Cao, P. Dagur, M. Driscoll,  A. Maritan, R. Maunu, J. P. McCoy Jr, R. B. Nussenblatt, H. N. Sen, and L. Wei, whose contributions to the Supercell/SVM approach~\cite{cand13} are extensively described in this review. J. C. was supported by NIH Award Number T32CA154274 from the National Cancer Institute. The content is solely the responsibility of the authors and does not necessarily represent the official views of the National Cancer Institute or the National Institutes of Health.

\end{document}